\documentclass[onecolumn,superscriptaddress,nobalancelastpage,12pt,pra,a4paper]{article}
\usepackage{graphicx} 
\usepackage{amsmath}
\usepackage{dcolumn}
\usepackage{bm}
\usepackage{verbatim}
\usepackage[USenglish]{babel}
\usepackage[dvipsnames]{xcolor}
\usepackage{braket}
\usepackage{dsfont}
\usepackage{physics}
\usepackage[backend=biber, style=numeric, sorting=none]{biblatex}
\usepackage{mathtools}
\usepackage{siunitx}[v=2]
\usepackage{subfigure}
\usepackage{subcaption}
\usepackage[T1]{fontenc}
\usepackage{authblk}
\usepackage[english, nameinlink]{cleveref}
\usepackage{soul}

\usepackage{titlesec}

\renewenvironment{psmallmatrix}
  {\left(\begin{smallmatrix}}
  {\end{smallmatrix}\right)}

\renewcommand{\d}{\mathrm{d}}

\newcommand{\I}{\mathcal{I}}
\newcommand{\kk}{\mathbf{k}}
\renewcommand{\qq}{\mathbf{q}}

\newcommand{\rr}{\mathbf{r}}
\newcommand{\pp}{\mathbf{p}}

\titleformat{\section}{\normalfont\large\bfseries}{}{0pt}{}
\titleformat{\subsection}{\normalfont}{}{0pt}{\ul}
\titlespacing{\subsection}{0pt}{\parskip}{4pt}
\titleformat{\subsubsection}{\normalfont\large}{}{0pt}{}
\titlespacing{\subsubsection}{0pt}{\parskip}{4pt}

\title{Vectorial field reconstruction without detecting the field}

\author{Jonas Vasikonis,$^{1,*}$ Sebastian Töpfer,$^{1}$ Satyajeet Patil,$^{1}$ Jorge Fuenzalida,$^{1,2}$ Markus Gräfe$^{1,3}$
\vspace{1em}
\\
\it $^{1}$Institute for Applied Physics, Technical University of Darmstadt, Otto-Berndt-Straße 3, 64287 Darmstadt, Germany\\
\it $^{2}$ICFO–Institut de Ciències Fotòniques, The Barcelona Institute of Science and Technology, Castelldefels 08860, Spain\\
\it $^{3}$Fraunhofer Institute for Applied Optics and Precision Engineering IOF, Albert-Einstein-Straße 7, 07745 Jena, Germany\\
$^{*}$email: jonas.vasikonis@tu-darmstadt.de 
}
\date{\today}

\begin{document}

\maketitle
\section{Abstract}
Vector beams, whose polarization varies across the transverse profile, are a central resource in structured-light optics and quantum photonics. Their characterization, however, becomes challenging when the field lies in a spectral region for which efficient spatially resolving detectors are unavailable. Here we demonstrate the spatially resolved reconstruction of an undetected vector beam by exploiting induced coherence in a nonlinear interferometer. In this effect, indistinguishability between two down-conversion pathways allows information encoded in an undetected field to be read out through interference of its detected partner. A telecom-wavelength idler field acquires a spatially varying polarization transformation but is never directly detected. Instead, its local polarization information is inferred from single-photon interference in the visible signal field, enabled by momentum correlations of the photon pair. Using phase-shifting and off-axis quantum holography with two polarization projections, we reconstruct the horizontal and vertical amplitudes and their relative phase across the beam profile, thereby recovering the full vectorial structure of the undetected field. We experimentally retrieve the polarization texture of an $m=2$ vector beam and compare multi-shot and single-shot reconstruction strategies. Our results extend imaging with undetected light from scalar objects to vectorial optical fields and open a route to polarization-sensitive sensing and state reconstruction in spectral regions that are difficult to access directly.
\section{From scalar undetected-light imaging to vectorial field reconstruction}

Vector beams, i.e.\ optical fields with a spatially varying polarization structure, have become an important resource in both classical and quantum optics~\cite{Sheppard:00,Maurer_2007,Zhan:09}. Their non-separable coupling of spatial and polarization degrees of freedom enables tightly focused field distributions and tailored light--matter interactions~\cite{Holleczek:11}, with applications ranging from lithography~\cite{Nivas2017}, confocal microscopy~\cite{10.1117/1.1382610}, optical trapping and efficient light--matter coupling~\cite{optical_tweeezer}, to laser material processing~\cite{meier_romano} and nanoscale probing~\cite{Banzer:10}. In the quantum regime, vector beams are equally attractive because they provide access to hybrid encoding across polarization and orbital angular momentum, and have been linked to squeezing, entanglement, and high-dimensional photonic state engineering~\cite{PhysRevLett.102.163602,PhysRevLett.106.060502}.

A central requirement for many of these applications is the ability to characterize the full transverse polarization structure of the field. This task becomes particularly challenging when the light of interest lies in spectral regions for which efficient, low-noise, spatially resolving detectors are unavailable, such as wavelength bands relevant for chemical-selective sensing and infrared imaging~\cite{Weissleder2001ACV}. In such cases, induced coherence offers an elegant alternative: when one output of two SPDC sources is mode-matched, the two emission pathways become indistinguishable, and single-photon interference appears in the partner field~\cite{PhysRevA.44.4614,PhysRevLett.67.318}. Information carried by the undetected field can then be inferred from interference of its correlated partner at a different wavelength.

Since its seminal demonstrations via nonlinear interferometers, induced coherence has become a powerful platform for quantum imaging and sensing~\cite{fuenzalida2024nonlinear}. Nonlinear interferometer enabled imaging with undetected light~\cite{Lemos_2014}, spectroscopy with visible-light detection of infrared features~\cite{Kalashnikov_2016}, control of polarization through quantum distinguishability~\cite{PhysRevA.95.033816}, tests of complementarity~\cite{PhysRevLett.114.053601}, two-colour interferometry~\cite{Hochrainer:17}, and access to biphoton correlations in single-photon interference~\cite{PhysRevA.96.013822,Brambila:25}.
More recently, quantum holographic approaches have shown that amplitude and phase information of an undetected field can be reconstructed through measurements on the detected partner~\cite{topfer2022quantum,pearce2024single,leon2024off,topfer2025synthetic}, and that these techniques also deliver resilience to noise~\cite{fuenzalida2023experimental}.  What has remained missing, however, is the extension of these concepts from scalar field reconstruction to the spatially resolved reconstruction of a vectorial optical field.

\begin{figure}[tb]
\centering
\includegraphics[width=\textwidth]{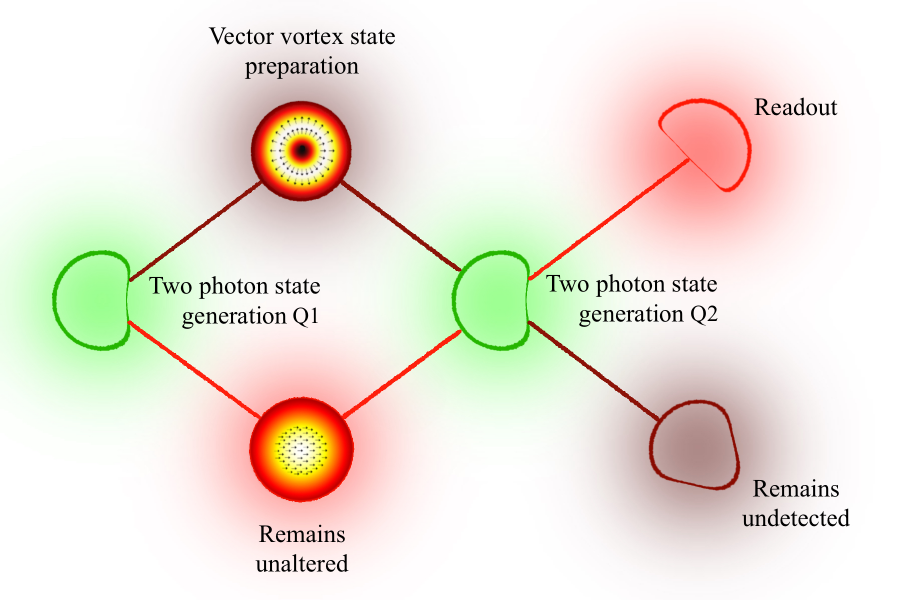}
\caption{
Principle of the characterization of undetected vector beams. A photon pair can be generated in either of two nonlinear sources. The vector state is prepared only in one idler arm, whereas the corresponding signal arm remains unaltered. After alignment of the idler modes, the two generation pathways become indistinguishable, so that induced coherence leads to interference in the detected signal field. The information encoded in the undetected idler field can thus be retrieved from the signal readout, while the idler field itself is never directly detected.
}
\label{fig:skizze_nli}
\end{figure}

Here, we fill this gap by combining induced coherence with quantum holography to reconstruct vector beams that remain undetected. As illustrated in Fig.~\ref{fig:skizze_nli}, a photon pair can be generated in either of two nonlinear sources (Q1 and Q2). While the signal field remains unchanged, the idler field from source 1 undergoes a spatially varying polarization transformation that prepares the vectorial state. Once this transformed idler mode is aligned with the idler mode from source 2, the two generation pathways become partially indistinguishable and interference appears in the detected signal field. Owing to the transverse-momentum correlations of the SPDC pair~\cite{walborn2010spatial,fuenzalida2022resolution}, this interference carries spatially resolved information about the undetected idler field and thereby gives access to its multimode vectorial structure. By recording the signal interference for two polarization projections and reconstructing the corresponding complex interference terms, we recover the local horizontal and vertical amplitudes together with their relative phase, and thus the full vectorial structure of the undetected field. Experimentally, we implement this scheme for a telecom-wavelength idler beam and reconstruct the polarization texture of an $m=2$ vector beam.

Cylindrically polarized vector beams are commonly generated using optical elements with an azimuthally varying fast axis~\cite{Yao:11,Zhan:09}. In our experiment, such a spatially varying polarization transformation is applied only in the undetected idler arm, while the signal arm serves solely as the readout channel. This extends imaging with undetected light from scalar objects to vectorial optical fields.

\section{Mapping undetected polarization onto visible-light interference}

\subsection{Induced coherence with space–polarization correlations}
\label{subsec: setup+theo}

The setup for the characterization of undetected vector beams is depicted in Fig.~\ref{fig:setup}. Photon pairs are generated in a coherent superposition from the two sources Q1 and Q2 via spontaneous parametric down-conversion (SPDC). Assuming identical nonlinear crystals, equal pump conditions, a paraxial regime, and a type-I non-degenerate SPDC process, the two-photon state can be written in the transverse-momentum basis as
\begin{align}
\label{Eq:source}
\ket{\Psi}=\frac{1}{\sqrt{2}}\int C(\qq_S, \qq_I)\, \left(\ket{H}_{\qq_{S1}}\ket{H}_{\qq_{I1}}+e^{i \theta }\ket{H}_{\qq_{S2}}\ket{H}_{\qq_{I2}} \right) \,\d \qq_S \, \d \qq_I .
\end{align}

Here, $\ket{H}_{\qq_{nl}}$ denotes a horizontally polarized single-photon state with transverse wave vector $\qq_{nl}$, where $n=S,I$ labels signal and idler, and $l=1,2$ labels the source. The phase $\theta$ accounts for the relative phase between the two emission pathways. The quantity $|C(\qq_S,\qq_I)|^2=P(\qq_S,\qq_I)$ is the joint transverse-momentum probability distribution of the SPDC pair~\cite{walborn2010spatial,fuenzalida2022resolution}. The state in Eq.~\eqref{Eq:source} therefore carries both polarization and transverse-momentum correlations.

Our aim is to reconstruct a spatially varying polarization state imprinted on the idler field emitted from Q1. For a given transverse-momentum component, we write the local idler polarization state as
\begin{align}
    \ket{H}_{\qq_{I1}} \rightarrow \ket{\psi}_{\qq_{I1}} =\alpha_{\qq_{I1}} \ket{H} + e^ {i \varphi_{\qq_{I1}} } \beta_{\qq_{I1}} \ket{V}.
\end{align}
Here, $\ket{V}$ denotes vertical polarization, with $\alpha^2+\beta^2=1$ and $\varphi\in[-\pi,\pi)$. This parametrization describes an arbitrary pure vector state across the transverse profile.

After this transformation, the idler field passes through HWP3, operated at either $0^\circ$ or $45^\circ$, before being overlapped with the idler mode associated with Q2. We first consider HWP3 at $0^\circ$, for which the idler polarization remains unchanged. In this configuration, the horizontally polarized idler component from Q1 is indistinguishable from the idler emitted by Q2, such that $\ket{H}_{\qq_{I1}}=\ket{H}_{\qq_{I2}}=\ket{H}_{\qq_I}$. The corresponding signal modes are recombined at a beam splitter and detected with a camera after spectral filtering.

When the idler modes are aligned, and the optical path differences lie within the coherence length of the SPDC photons~\cite{PhysRevA.44.4614,PhysRevLett.67.318}, the which-source information is erased, and single-photon interference appears in the signal beam. For HWP3 set to $0^\circ$, the detected signal intensity is
\begin{align}
    \I_{\qq_{S}} \approx  1+  \alpha_{\qq_{I}} \cos(\theta').
    \label{Eq:intsignal1}
\end{align}
Here, $\theta'$ contains the accumulated interferometric phase, including the relative path phase and the spatially invariant phase offset of the interferometer. Each camera pixel corresponds to a transverse-momentum component $\pp_S=\hbar \qq_S$ of the signal field. Through the momentum correlations encoded in $C(\qq_S,\qq_I)$, this detected component carries information about the corresponding idler component.

In the second configuration, HWP3 is set to $45^\circ$, so that the interference becomes sensitive to the vertical component of the idler polarization. The signal intensity then reads
\begin{align}
    \I_{\qq_{S}} \approx  1+  \beta_{\qq_{I}} \cos(\theta'+ \varphi_{\qq_{I}} ).
    \label{Eq:intsignal2}
\end{align}
Equations~\eqref{Eq:intsignal1} and~\eqref{Eq:intsignal2} show that the interference of the detected signal field provides access to the local polarization parameters $\alpha_{\qq_I}$, $\beta_{\qq_I}$, and $\varphi_{\qq_I}$ of the undetected idler field. In this way, the full vectorial structure of the undetected field can be reconstructed without direct idler detection.

\subsection{Imaging system and vector-beam preparation}

The creation of vector beams requires employing spatial modes to encode the polarization information. In our case, the vector beam is prepared using the transverse momentum of the idler photon, $\qq_I$. The vector beam's information is later recovered using the transverse momentum of the signal photon, $\qq_S$, through the connection between these two momenta in Eq.~\eqref{Eq:source}. The creation of the vector beam and its later detection is explained next.

\begin{figure}[tb]
\centering
\includegraphics[width=\textwidth]{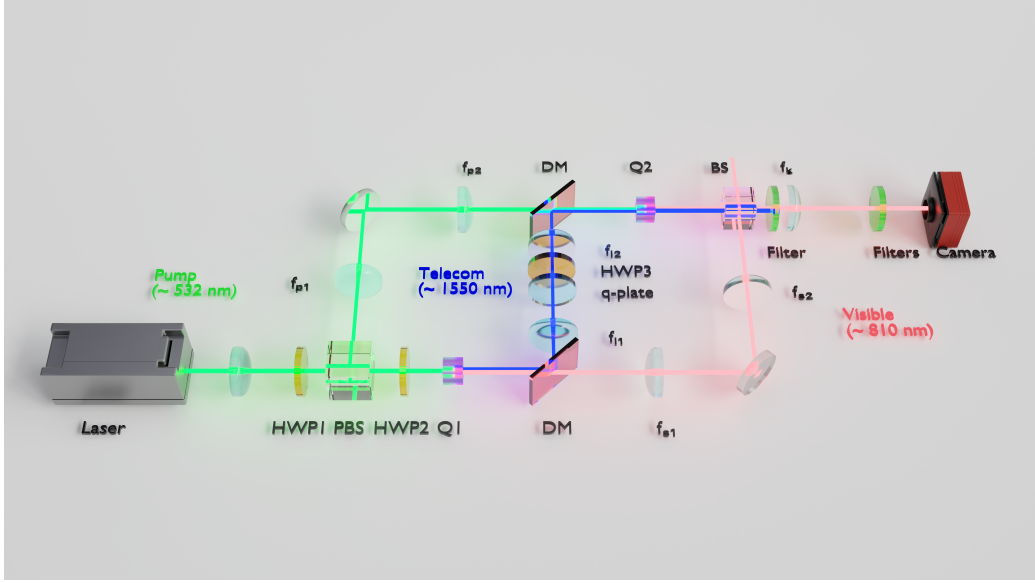}
\caption{Experimental setup for vectorial field reconstruction with undetected light. A continuous-wave single-mode pump laser at $\SI{532}{\nano\meter}$ pumps two nonlinear PPLN crystals, Q1 and Q2. The idler field at $\SI{1550}{\nano\meter}$ from Q1 is structured by a q-plate and a half-wave plate (HWP3), overlapped with the idler mode at Q2, and remains undetected. The corresponding signal fields at $\SI{810}{\nano\meter}$ are recombined at a beam splitter and detected on an sCMOS camera after spectral filtering.}
\label{fig:setup}
\end{figure}

Let us consider a plane wave of the idler beam, $\kk_{I}$, emitted at the source Q1, see Fig.~\ref{fig:setup}. The lens f$_{I1}$ of focal distance $f$ located at a distance $f$ from Q1, performs a Fourier transform on the idler beam at its back focal plane, producing the far-field plane. One point on this plane represents a transverse momentum component, with $\pp_I=\hbar\qq_{I}$, i.e., the initial plane wave is focused to a point represented by $f_{I1} \, \lambda_{I} \, \qq_I/ 2\pi$; see Refs.~\cite{theorymayukh,fuenzalida2022resolution} for more details. 
In the far field plane, a vortex wave plate is placed to change the polarization at different transverse positions of the idler photon. Employing Jones matrix formalism~\cite{qplate_matrix0}, the transformation $\textbf{M}$ of the m-plate to each transverse point is given by $\textbf{M}=\textbf{R}(-\theta) \cdot \sigma_z \cdot \textbf{R}(\theta)$, where $\textbf{R}$ is the standard 2$\times$2 rotation matrix, and $\sigma_z= \begin{psmallmatrix} 1 & 0 \\ 0 & -1 \end{psmallmatrix}$. Hence 
\begin{align}
    \textbf{M}= \begin{pmatrix}  \cos(2 \phi) &  \sin(2 \phi) \\  \sin(2 \phi) & - \cos(2 \phi) \end{pmatrix}
\end{align}
where $\phi(r,\gamma)=m \gamma/2+ \phi_0 $. m is the order of the plate, $\gamma$ is the azimuthal angle, and $\phi_0$ is the orientation of the fast axis at $\gamma=0.$
After the idler photon acquires the vector beam form, it passes through a second lens, f$_{I2}$, that performs a second Fourier transform on it. This way, the idler photon is imaged from Q1 to Q2 using a 4f system (with lenses f$_{I1}$ and f$_{I2}$). This optical system images the transformed idler mode from Q1 onto the mode that overlaps with the idler mode of Q2. Likewise, the signal beam emitted in Q1 is imaged with a 4f system (with lenses f$_{S1}$ and f$_{S2}$), and subsequently, a third lens, f$_k$, produces a Fourier transform on it, such that the camera detects its far-field plane. From source Q2, the signal beam is also Fourier-transformed by the lens f$_k$, and the camera detects its far-field too. In this way, one plane wave of the signal photon is focused to a point at the camera plane represented by $f_{k} \, \lambda_{S} \, \qq_S/ 2\pi$. Due to spatial correlations of SPDC photons, and the effect of induced coherence,  the vector beam information is transferred from the idler to the signal through the photon pair momentum correlation.


\subsection{Vectorial reconstruction protocol}
\label{subsec:holography}
To reconstruct the local polarization state of the undetected idler field, we record the interference of the detected signal field for two polarization projections, corresponding to HWP3 angles of $0^\circ$ and $45^\circ$. From these two measurements, we retrieve the local coefficients $\alpha_{\qq_I}$ and $\beta_{\qq_I}$ together with the relative phase $\varphi_{\qq_I}$.

We implement this reconstruction using two complementary quantum holographic approaches. In phase-shifting digital holography (PSDH), four interferograms are recorded for relative phase steps of $0$, $\pi/2$, $\pi$, and $3\pi/2$, allowing the complex interference term to be extracted pixel by pixel~\cite{yamaguchi1997phase,topfer2022quantum}. In off-axis holography (OAH), a small angular tilt between the interfering fields shifts the interference term in Fourier space, enabling single-shot retrieval of the complex field for each polarization projection~\cite{Cuche:00,Verrier:11,pearce2024single,leon2024off,topfer2025synthetic}.

In the following, we first use PSDH as a multi-frame reconstruction strategy and then compare it with OAH as a single-frame alternative for recovering the vectorial structure of the undetected field.
\section{Multi-shot and single-shot reconstruction of an undetected vector beam}
\label{sec:result}
\subsection{Multi-frame Jones reconstruction by phase-shifting holography}

\begin{figure}[htbp]
    \centering   
    \includegraphics[width=\textwidth]{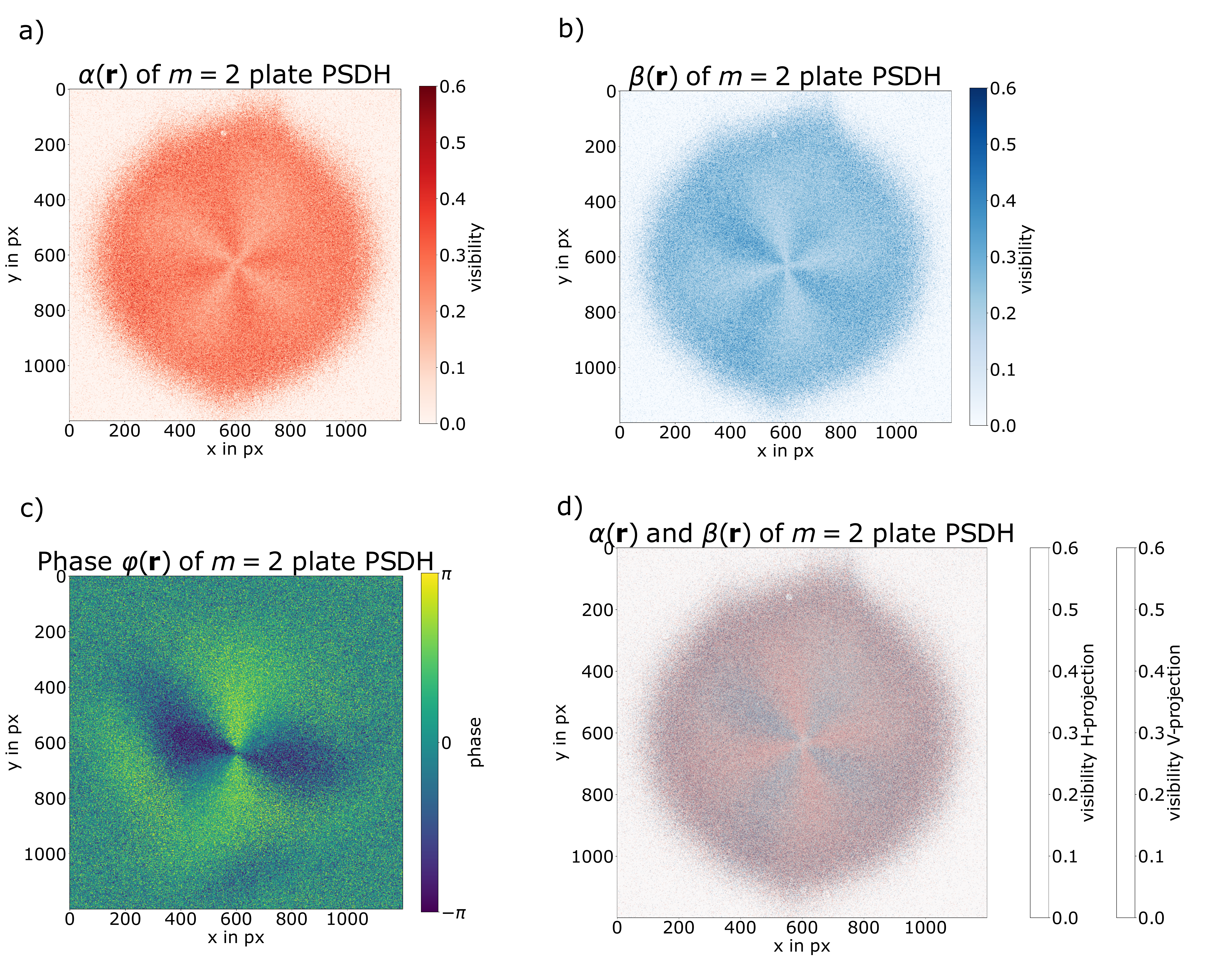}
    
    
    \caption{Phase-shifting holography reconstruction of the local polarization amplitudes (a,b,d) and phase (c)  of the undetected vector field generated with a vortex wave plate of topological charge $m=2$. While Image a) shows the reconstructed H-projection $\alpha(\rr)$, b) shows the reconstructed V-projection $\beta(\rr)$. Both exhibit a visibility up to $0.6$. In c) the phase $\varphi(\rr)$ is shown. This function extends from $-\pi$ to $\pi$. The combined image of both projections is depicted in d)}
    \label{fig:vis_alpha_beta}
\end{figure}

We first reconstruct the undetected vector field using PSDH. The retrieved spatial maps for the amplitudes $\alpha$ and $\beta$ are shown in Fig.~\ref{fig:vis_alpha_beta} (a) \& (b). These maps demonstrate that the horizontal and vertical components of the undetected idler field can be reconstructed across the transverse profile from measurements performed solely on the detected signal field. The corresponding phase map $\varphi$ is shown in Fig.~\ref{fig:vis_alpha_beta}(c). Together, the reconstructed amplitudes yield the combined representation shown in Fig.~\ref{fig:vis_alpha_beta}(d), which reproduces the polarization texture expected for the $m=2$ vector beam generated by the vortex wave plate. 

These measurements establish that the full local Jones information of the undetected field can be recovered from the interference of the detected field. In particular, PSDH provides direct access to the two polarization amplitudes and their relative phase in a conceptually simple and experimentally robust way, at the cost of a multi-frame acquisition.

\subsection{Single-shot vector-field retrieval by off-axis holography}

We next implement OAH as a single-shot alternative to PSDH. In this case, a small angular offset between the interfering signal fields produces spatial carrier fringes, allowing the complex interference term to be isolated in Fourier space. Only one interferogram is required for each polarization projection, so that a complete reconstruction is obtained from two images in total.

\begin{figure}[htbp]
    \centering   
    \includegraphics[width=\textwidth]{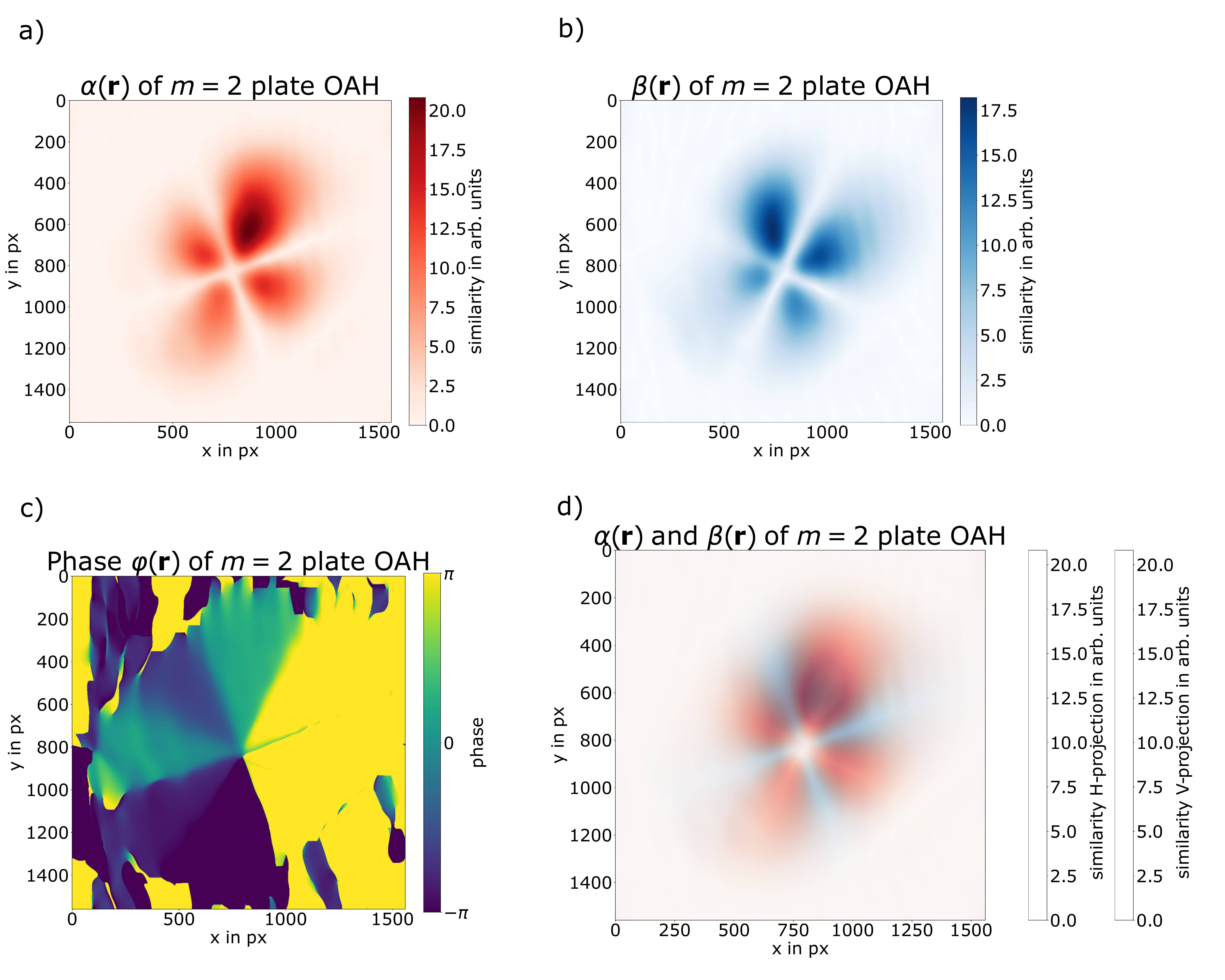}
    \caption{ Off-axis holography reconstruction of the local polarization amplitudes (a,b,d) and phase (c) of the undetected vector field generated with a vortex wave plate of topological charge $m=2$.While Image a) shows the reconstructed H-projection $\alpha(\rr)$, b) shows the reconstructed V-projection $\beta(\rr)$. In c) the phase $\varphi(\rr)$ is shown. This function extends from $-\pi$ to $\pi$.  The combined image of both projections is depicted in d)}
    \label{fig:vis_alpha_beta_off}
\end{figure}

The reconstructed amplitude maps obtained with OAH are shown in Fig.~\ref{fig:vis_alpha_beta_off}(a) \& (b). Compared with the PSDH reconstruction, the OAH amplitude retrieval exhibits a cleaner background and an increased contrast of the spatial features. The corresponding phase reconstruction is shown in Fig.~\ref{fig:vis_alpha_beta_off}(c). The phase retains the expected spatial variation, although the recovered map includes a partially obscured region, which is also reflected in the combined state shown in Fig.~\ref{fig:vis_alpha_beta_off}(d). Despite this local degradation, the overall vectorial structure of the undetected field remains clearly identifiable.

The OAH results therefore highlight the trade-off between the two reconstruction strategies. PSDH provides a direct multi-frame route to the complex field, whereas OAH substantially reduces the number of required images and is therefore attractive for faster acquisitions and for measurements that are more sensitive to phase drifts. In our implementation, OAH improves the amplitude reconstruction while introducing a localized imperfection in the recovered phase.

\begin{figure}[htbp]
    \centering   
    \includegraphics[width=\textwidth]{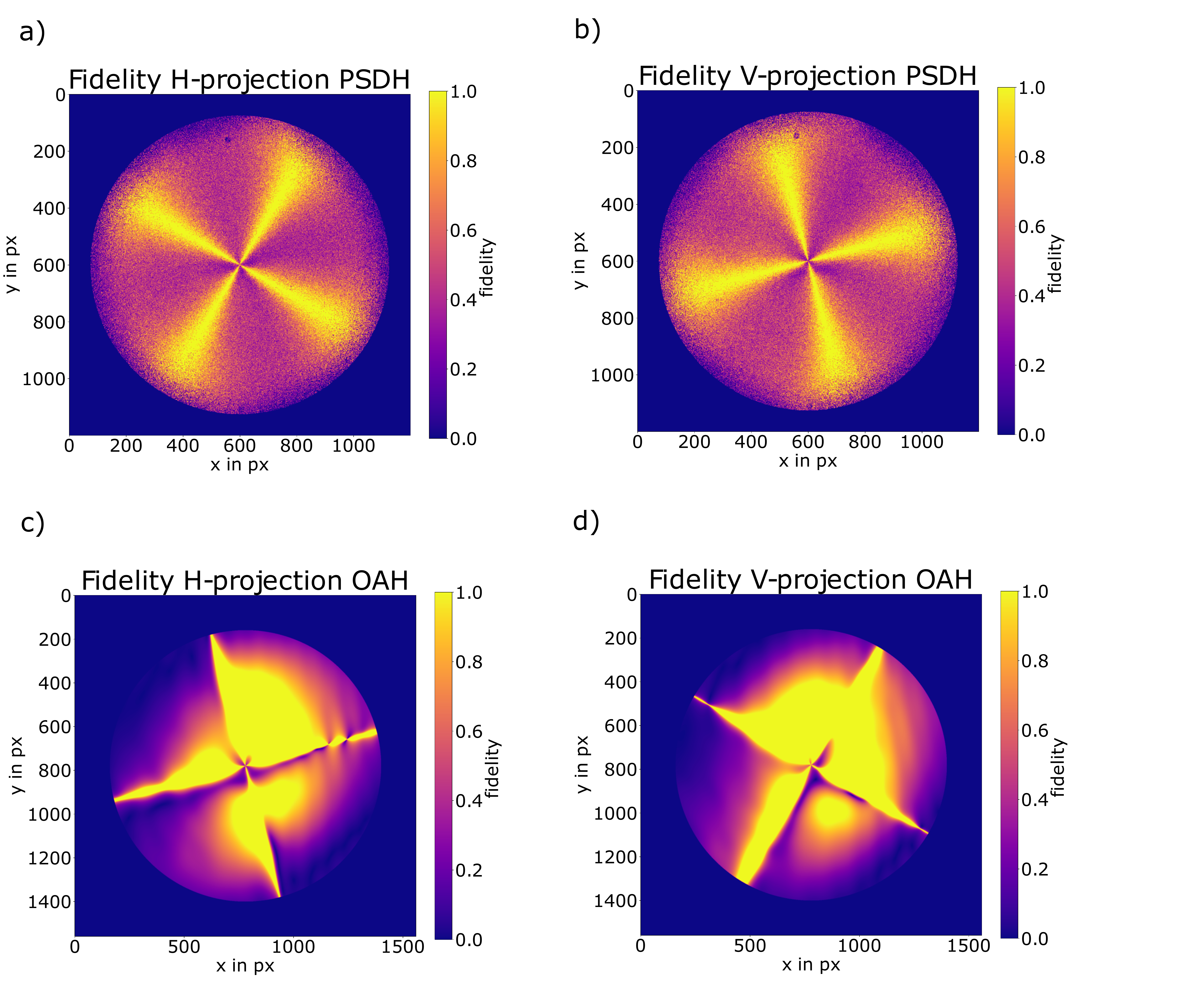}
    \caption{Fidelity for reconstructed polariaztion amplitudes of phase-shifting digital holography (a,b) and off-axis holography (c,d).The fidelity of the H-projection is shown in Figures (a) and (c).The fidelity of the V-projection is displayed in (b) and (d).(a) and (b) demonstrate the fidelity for PSDH, while (c) and (d) demonstrate the fidelity for OAH.}
    \label{fig:fidelity}
\end{figure}

%

%

To compare both reconstruction strategies quantitatively, we evaluate the retrieved amplitude quality, the background level, and the phase quality in the reconstructed images. Relative to PSDH, OAH yields a $\SI{22.9}{\%}$ increase in amplitude  reconstruction quality and a a $\SI{5.43}{\%}$ reduction of the background level within the region of interest. Additionally, OAH has been shown to enhance the signal-to-noise ratio (SNR) by $\SI{14.81}{\%}$ in comparison with PSDH. However, the phase reconstruction exhibits a localized degradation over approximately $\SI{18.97}{\%}$ of the reconstructed area. The remaining field of view retains the expected spatial phase variation and reproduces the overall vectorial structure of the undetected field. These values confirm that OAH provides a more efficient amplitude readout, whereas PSDH remains advantageous when uniform phase fidelity across the full beam profile is required.


\section{Capabilities and limits of vectorial undetected-light imaging}
\label{sec:discussion}

The present experiment extends induced-coherence imaging into a regime in which the relevant information is not only scalar, but intrinsically vectorial. In earlier demonstrations with undetected light, the reconstructed quantities were typically associated with transmission, phase, or related scalar field properties. Here, by contrast, the information of interest is encoded in the local polarization structure of the undetected field. The ability to recover $\alpha_{\qq_I}$, $\beta_{\qq_I}$, and $\varphi_{\qq_I}$ across the beam profile therefore shows that induced coherence can be used to access a genuinely higher level of optical structure than in standard undetected-light imaging schemes.

This is conceptually significant because polarization is not simply an additional observable, but part of the internal structure of the field. In the present scheme, that internal structure is mapped onto a detected interference signal through the interplay of three ingredients: induced coherence between the two nonlinear sources, transverse-momentum correlations of the SPDC pair, and a spatially varying polarization transformation in the undetected arm. The experiment thus links structured-light optics and nonlinear interferometry in a particularly direct way. Rather than probing whether an undetected object modifies a scalar field, the interferometer becomes sensitive to how the undetected field itself is structured locally in polarization.

From this perspective, the key result is not only that a particular $m=2$ vector beam can be reconstructed, but that the nonlinear interferometer acts as a wavelength-translating analyzer of vectorial field structure. The measurement is carried out entirely in the visible arm, while the relevant polarization texture is imposed at 1550~nm. This separation between preparation and detection is especially attractive in spectral regions where direct polarization-resolved imaging is technically difficult, noisy, or expensive. The present proof-of-principle therefore points beyond a specific beam class and towards a more general strategy for vectorial sensing with undetected light.

The comparison between PSDH and OAH also clarifies the practical trade-off between both reconstruction strategies. Quantitatively, OAH provides the higher amplitude contrast and the lower background level, whereas PSDH yields the more homogeneous phase reconstruction across the full beam profile. At the same time, the OAH phase map shows a localized degradation, indicating that the gain in acquisition efficiency comes at the price of increased sensitivity to Fourier-space filtering and alignment conditions. The two methods should therefore not be viewed simply as a slower and a faster implementation of the same protocol, but as complementary operating modes with different strengths: OAH is advantageous for rapid, high-contrast amplitude retrieval, whereas PSDH remains the more robust option when phase fidelity across the entire field is the critical figure of merit.

An important feature of vector beams is the polarization singularity at the beam centre. In the present data, the global $m=2$ polarization texture is clearly recovered, whereas the singular core itself is not sharply resolved. This is not unexpected: the local phase becomes ill-defined in regions where the reconstructed amplitudes approach zero, and the finite spatial resolution, limited visibility, and Fourier-space filtering further reduce the fidelity of the reconstruction near the beam centre. The present experiment therefore resolves the surrounding vectorial structure reliably, while the singular point itself remains the most demanding region for reconstruction. A more complete treatment of polarization singularities will require improved spatial resolution and a reconstruction strategy tailored to low-signal regions.

The current implementation also makes clear where the method remains limited. The reconstruction is formulated for locally pure polarization states and demonstrated for a known vector field generated by a vortex wave plate. The present work therefore establishes the principle of vectorial field reconstruction with undetected light, but it does not yet constitute a general polarization tomography framework for arbitrary partially polarized or mixed vector fields. Extending the method in that direction would require additional projections and a more general reconstruction formalism, potentially involving Stokes-parameter retrieval or overcomplete measurement sets (cf. Refs.~\cite{fuenzalida2023quantum,kysela2025visibility}).

More specifically, the present results show that induced-coherence imaging can be extended from scalar to vectorial field reconstruction. The novelty is therefore not generic access to arbitrary undetected states, but the ability to recover a spatially varying polarization structure at an undetected wavelength through measurements performed entirely in the detected arm. This establishes a concrete route towards polarization-sensitive reconstruction schemes with undetected light.
\section{Experimental implementation and holographic retrieval}
\label{sec:methods}
\subsection{Nonlinear interferometer for visible readout of a telecom idler}
\label{subsec:exp_setup}
A continuous-wave laser at 532~nm pumps two identical nonlinear sources Q1 and Q2, which generate signal and idler fields at 810~nm and 1550~nm, respectively, via type-0 non-degenerate SPDC. The nonlinear crystals have dimensions of $1\times1\times2~\mathrm{mm}^3$ (height $\times$ width $\times$ length). Signal and idler emitted from Q1 are separated by the dichroic mirror DM1.

The imaging systems are implemented with lenses of focal lengths $f_{P1}=f_{P2}=f_{I1}=f_{I2}=f_{S1}=f_{S2}=75~\mathrm{mm}$ and $f_k=150~\mathrm{mm}$. A vortex wave plate of topological charge $m=2$ is placed in the idler path. The second dichroic mirror DM2 recombines the idler field with the pump before impinging Q2. Owing to the imaging conditions, the SPDC processes in Q1 and Q2 have identical spatial and spectral characteristics.

The signal fields emitted from Q1 and Q2 are recombined at a beam splitter and detected on an sCMOS camera. Two spectral filters centred at $810\pm10~\mathrm{nm}$ suppress background light and reject the idler field. The camera records one image per second. For PSDH, a piezoelectric actuator with nanometre-scale resolution is used to introduce the required phase steps. For OAH, no piezo scan is required; instead, a small angular offset between the interfering signal fields produces the spatial carrier needed for Fourier-domain reconstruction.

\begin{figure}[htbp]
    \centering   
    \includegraphics[width=\textwidth]{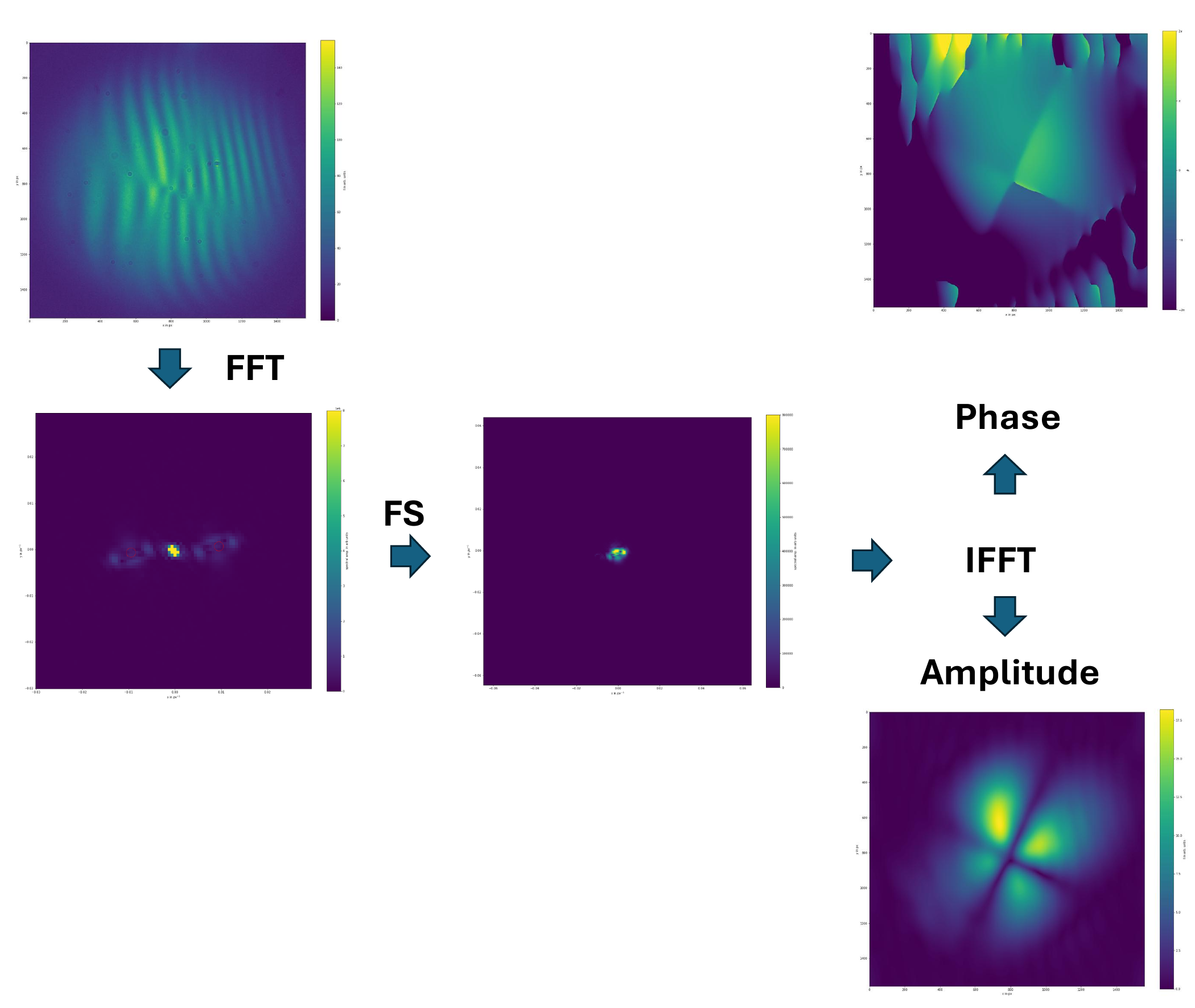}
    \caption{Schematics for OAH. The recorded image is converted to Fourier space via a fast Fourier transformation (FFT). In Fourier space, the first order is separated and can be suppressed by a first-order shift (FS). After an inverse fast Fourier transformation (IFFT), the complex field is obtained, from which the amplitude (modulus) and phase (argument) can be derived.}
    \label{fig:expl}
\end{figure}

\subsection{Phase-stepped extraction of amplitude and relative phase}
\label{subsec:method_psdh}

For PSDH, the interferometric phase $\theta'$ is stepped through the values $0$, $\pi/2$, $\pi$, and $3\pi/2$. From the resulting four interferograms, the local interference amplitude is obtained as~\cite{yamaguchi1997phase,topfer2022quantum}
\begin{align}
    A_{\qq_{I}}=
    2\,
    \frac{
    \sqrt{
    \left[\I_{\qq_{S}}(3\pi/2)-\I_{\qq_{S}}(\pi/2)\right]^2+
    \left[\I_{\qq_{S}}(0)-\I_{\qq_{S}}(\pi)\right]^2
    }
    }{
    \I_{\qq_{S}}(0)+\I_{\qq_{S}}(\pi/2)+\I_{\qq_{S}}(\pi)+\I_{\qq_{S}}(3\pi/2)
    }.
\end{align}
For HWP3 set to $0^\circ$, this amplitude corresponds to $\alpha_{\qq_I}$, whereas for HWP3 set to $45^\circ$ it corresponds to $\beta_{\qq_I}$.

The relative phase is reconstructed from the same four measurements as
\begin{align}
\varphi_{\qq_{I}}=
\arctan\left(
\frac{\I_{\qq_{S}}(3\pi/2)-\I_{\qq_{S}}(\pi/2)}
{\I_{\qq_{S}}(0)-\I_{\qq_{S}}(\pi)}
\right).
\end{align}
A complete PSDH reconstruction therefore requires four interferograms for each polarization projection and eight images in total.For further details and information, please refer to Ref \cite{topfer2022quantum}. 

\subsection{Fourier-domain retrieval from spatial carrier fringes}
\label{subsec:method_oah}
For OAH, the two interfering signal fields are aligned with a small relative angle, such that the recorded camera intensity takes the standard holographic form~\cite{Cuche:00,Verrier:11}
\begin{align}
    \label{eq:oah}
    I(\rr)=\left|E_O(\rr)+E_R(\rr)\right|^2
    =\left|E_O(\rr)\right|^2+\left|E_R(\rr)\right|^2
    +E_O^*(\rr)E_R(\rr)+E_R^*(\rr)E_O(\rr),
\end{align}
where $E_O(\rr)$ and $E_R(\rr)$ denote the object and reference fields, respectively.

If the reference field carries a transverse wave vector $\kk$, the Fourier transform of Eq.~\eqref{eq:oah} contains two displaced first-order terms~\cite{leon2024off}
\begin{align}
    \label{eq:ft_oah}
    \tilde{I}(\qq)=DC(\qq)+\tilde{E}_O^*(\qq)\otimes\delta(\qq-\kk)+\tilde{E}_O(\qq)\otimes\delta(\qq+\kk),
\end{align}
where the first term contains the zero-order contribution and the latter two correspond to the shifted first-order terms. One of these first-order terms is isolated in Fourier space, shifted to the origin, and inverse Fourier transformed to recover the complex field. Its modulus yields the local amplitude, and its argument yields the local phase~\cite{Szadowiak:2021lpu,Sanchez_Ortiga:2014AO}.

In our implementation, one interferogram is recorded for each polarization projection. OAH therefore reduces the required data set to two images in total, at the expense of the additional Fourier-space filtering step shown schematically in Fig.~\ref{fig:expl}.
\pagebreak
\printbibliography[title={Literature},heading=bibnumbered]

@article{Lemos_2014,
   title={Quantum imaging with undetected photons},
   volume={512},
   ISSN={1476-4687},
   url={http://dx.doi.org/10.1038/nature13586},
   DOI={10.1038/nature13586},
   number={7515},
   journal={Nature},
   publisher={Springer Science and Business Media LLC},
   author={Lemos, Gabriela Barreto and Borish, Victoria and Cole, Garrett D. and Ramelow, Sven and Lapkiewicz, Radek and Zeilinger, Anton},
   year={2014},
   month={08}, 
   pages={409–412}
}

@article{Kalashnikov_2016,
   title={Infrared spectroscopy with visible light},
   volume={10},
   ISSN={1749-4893},
   url={http://dx.doi.org/10.1038/nphoton.2015.252},
   DOI={10.1038/nphoton.2015.252},
   number={2},
   journal={Nature Photonics},
   publisher={Springer Science and Business Media LLC},
   author={Kalashnikov, Dmitry A. and Paterova, Anna V. and Kulik, Sergei P. and Krivitsky, Leonid A.},
   year={2016},
   month={01}, 
   pages={98–101} 
}

@article{PhysRevA.95.033816,
  title = {Partial polarization by quantum distinguishability},
  author = {Lahiri, Mayukh and Hochrainer, Armin and Lapkiewicz, Radek and Lemos, Gabriela Barreto and Zeilinger, Anton},
  journal = {Phys. Rev. A},
  volume = {95},
  issue = {3},
  pages = {033816},
  numpages = {6},
  year = {2017},
  month = {03},
  publisher = {American Physical Society},
  doi = {10.1103/PhysRevA.95.033816},
  url = {https://link.aps.org/doi/10.1103/PhysRevA.95.033816}
}

@article{PhysRevLett.114.053601,
  title = {Induced Coherence, Vacuum Fields, and Complementarity in Biphoton Generation},
  author = {Heuer, A. and Menzel, R. and Milonni, P. W.},
  journal = {Phys. Rev. Lett.},
  volume = {114},
  issue = {5},
  pages = {053601},
  numpages = {5},
  year = {2015},
  month = {02},
  publisher = {American Physical Society},
  doi = {10.1103/PhysRevLett.114.053601},
  url = {https://link.aps.org/doi/10.1103/PhysRevLett.114.053601}
}

@article{Hochrainer:17,
author = {Armin Hochrainer and Mayukh Lahiri and Radek Lapkiewicz and Gabriela B. Lemos and Anton Zeilinger},
journal = {Optica},
number = {3},
pages = {341--344},
publisher = {Optica Publishing Group},
title = {Interference fringes controlled by noninterfering photons},
volume = {4},
month = {03},
year = {2017},
url = {https://opg.optica.org/optica/abstract.cfm?URI=optica-4-3-341},
doi = {10.1364/OPTICA.4.000341},
}

@article{PhysRevA.96.013822,
  title = {Twin-photon correlations in single-photon interference},
  author = {Lahiri, Mayukh and Hochrainer, Armin and Lapkiewicz, Radek and Lemos, Gabriela Barreto and Zeilinger, Anton},
  journal = {Phys. Rev. A},
  volume = {96},
  issue = {1},
  pages = {013822},
  numpages = {9},
  year = {2017},
  month = {07},
  publisher = {American Physical Society},
  doi = {10.1103/PhysRevA.96.013822},
  url = {https://link.aps.org/doi/10.1103/PhysRevA.96.013822}
}

@article{Weissleder2001ACV,
  title={A clearer vision for in vivo imaging},
  author={Ralph Weissleder},
  journal={Nature Biotechnology},
  year={2001},
  volume={19},
  pages={316-317},
  url={https://api.semanticscholar.org/CorpusID:2177544}
}

@article{fuenzalida2023quantum,
  title = {Quantum state tomography of undetected photons},
  author = {Fuenzalida, Jorge and Kysela, Jaroslav and Dovzhik, Krishna and Lemos, Gabriela Barreto and Hochrainer, Armin and Lahiri, Mayukh and Zeilinger, Anton},
  journal = {Phys. Rev. A},
  volume = {109},
  issue = {2},
  pages = {022413},
  numpages = {8},
  year = {2024},
  month = {02},
  publisher = {American Physical Society},
  doi = {10.1103/PhysRevA.109.022413},
  url = {https://link.aps.org/doi/10.1103/PhysRevA.109.022413}
}

@article{theorymayukh,
  title = {Theory of quantum imaging with undetected photons},
  author = {Lahiri, Mayukh and Lapkiewicz, Radek and Lemos, Gabriela Barreto and Zeilinger, Anton},
  journal = {Phys. Rev. A},
  volume = {92},
  issue = {1},
  pages = {013832},
  numpages = {8},
  year = {2015},
  month = {07},
  publisher = {American Physical Society},
  doi = {10.1103/PhysRevA.92.013832},
  url = {https://link.aps.org/doi/10.1103/PhysRevA.92.013832}
}

@article{kysela2025visibility,
  title = {Visibility Stokes parameters as a foundation for quantum information science with undetected photons},
  author = {Kysela, Jaroslav and Gr\"afe, Markus and Fuenzalida, Jorge},
  journal = {Phys. Rev. Res.},
  volume = {7},
  issue = {2},
  pages = {023133},
  numpages = {14},
  year = {2025},
  month = {05},
  publisher = {American Physical Society},
  doi = {10.1103/PhysRevResearch.7.023133},
  url = {https://link.aps.org/doi/10.1103/PhysRevResearch.7.023133}
}

@article{Zhan:09,
author = {Qiwen Zhan},
journal = {Adv. Opt. Photon.},
number = {1},
pages = {1--57},
publisher = {Optica Publishing Group},
title = {Cylindrical vector beams: from mathematical concepts to               applications},
volume = {1},
month = {01},
year = {2009},
url = {https://opg.optica.org/aop/abstract.cfm?URI=aop-1-1-1},
doi = {10.1364/AOP.1.000001},
}

@article{PhysRevLett.106.060502,
  title = {Entangling Different Degrees of Freedom by Quadrature Squeezing Cylindrically Polarized Modes},
  author = {Gabriel, C. and Aiello, A. and Zhong, W. and Euser, T. G. and Joly, N. Y. and Banzer, P. and F\"ortsch, M. and Elser, D. and Andersen, U. L. and Marquardt, Ch. and Russell, P. St. J. and Leuchs, G.},
  journal = {Phys. Rev. Lett.},
  volume = {106},
  issue = {6},
  pages = {060502},
  numpages = {4},
  year = {2011},
  month = {02},
  publisher = {American Physical Society},
  doi = {10.1103/PhysRevLett.106.060502},
  url = {https://link.aps.org/doi/10.1103/PhysRevLett.106.060502}
}

@article{PhysRevLett.67.318,
  title = {Induced coherence and indistinguishability in optical interference},
  author = {Zou, X. Y. and Wang, L. J. and Mandel, L.},
  journal = {Phys. Rev. Lett.},
  volume = {67},
  issue = {3},
  pages = {318--321},
  numpages = {0},
  year = {1991},
  month = {07},
  publisher = {American Physical Society},
  doi = {10.1103/PhysRevLett.67.318},
  url = {https://link.aps.org/doi/10.1103/PhysRevLett.67.318}
}

@article{PhysRevA.44.4614,
  title = {Induced coherence without induced emission},
  author = {Wang, L. J. and Zou, X. Y. and Mandel, L.},
  journal = {Phys. Rev. A},
  volume = {44},
  issue = {7},
  pages = {4614--4622},
  numpages = {0},
  year = {1991},
  month = {10},
  publisher = {American Physical Society},
  doi = {10.1103/PhysRevA.44.4614},
  url = {https://link.aps.org/doi/10.1103/PhysRevA.44.4614}
}

@article{Holleczek:11,
author = {Annemarie Holleczek and Andrea Aiello and Christian Gabriel and Christoph Marquardt and Gerd Leuchs},
journal = {Opt. Express},
number = {10},
pages = {9714--9736},
publisher = {Optica Publishing Group},
title = {Classical and quantum properties of cylindrically polarized states of light},
volume = {19},
month = {05},
year = {2011},
url = {https://opg.optica.org/oe/abstract.cfm?URI=oe-19-10-9714},
doi = {10.1364/OE.19.009714},

}

@article{10.1117/1.1382610,
author = {Nils Huse and Andreas Schoenle and Stefan W. Hell},
title = {{Z-polarized confocal microscopy}},
volume = {6},
journal = {Journal of Biomedical Optics},
number = {3},
publisher = {SPIE},
pages = {273 -- 276},
year = {2001},
doi = {10.1117/1.1382610},
URL = {https://doi.org/10.1117/1.1382610}
}

@ARTICLE{optical_tweeezer,
       author = {{Sondermann}, M. and {Maiwald}, R. and {Konermann}, H. and {Lindlein}, N. and {Peschel}, U. and {Leuchs}, G.},
        title = "{Design of a mode converter for efficient light-atom coupling in free space}",
      journal = {Applied Physics B: Lasers and Optics},
         year = 2007,
        month = 12,
       volume = {89},
       number = {4},
        pages = {489-492},
          doi = {10.1007/s00340-007-2859-4},
archivePrefix = {arXiv},
       eprint = {0708.0772},
 primaryClass = {quant-ph},
       adsurl = {https://ui.adsabs.harvard.edu/abs/2007ApPhB..89..489S}
}

@article{meier_romano,
author = {Meier, M. and Romano, Valerio and Feurer, T.},
year = {2007},
month = {03},
pages = {329-334},
title = {Material processing with pulsed radially and azimuthally polarized laser radiation},
volume = {86},
journal = {Applied Physics A},
doi = {10.1007/s00339-006-3784-9}
}

@article{Banzer:10,
author = {P. Banzer and U. Peschel and S. Quabis and G. Leuchs},
journal = {Opt. Express},
number = {10},
pages = {10905--10923},
publisher = {Optica Publishing Group},
title = {On the experimental investigation of the electric and magnetic response of a single nano-structure},
volume = {18},
month = {05},
year = {2010},
url = {https://opg.optica.org/oe/abstract.cfm?URI=oe-18-10-10905},
doi = {10.1364/OE.18.010905},

}

@article{PhysRevLett.102.163602,
  title = {Continuous Variable Entanglement and Squeezing of Orbital Angular Momentum States},
  author = {Lassen, M. and Leuchs, G. and Andersen, U. L.},
  journal = {Phys. Rev. Lett.},
  volume = {102},
  issue = {16},
  pages = {163602},
  numpages = {4},
  year = {2009},
  month = {04},
  publisher = {American Physical Society},
  doi = {10.1103/PhysRevLett.102.163602},
  url = {https://link.aps.org/doi/10.1103/PhysRevLett.102.163602}
}

@article{Maurer_2007,
doi = {10.1088/1367-2630/9/3/078},
url = {https://dx.doi.org/10.1088/1367-2630/9/3/078},
year = {2007},
month = {03},
publisher = {},
volume = {9},
number = {3},
pages = {78},
author = {Christian Maurer and Alexander Jesacher and Severin Fürhapter and Stefan Bernet and Monika Ritsch-Marte},
title = {Tailoring of arbitrary optical vector beams},
journal = {New Journal of Physics},

}

@article{Sheppard:00,
author = {Colin J. R. Sheppard},
journal = {J. Opt. Soc. Am. A},
number = {2},
pages = {335--341},
publisher = {Optica Publishing Group},
title = {Polarization of almost-plane waves},
volume = {17},
month = {02},
year = {2000},
url = {https://opg.optica.org/josaa/abstract.cfm?URI=josaa-17-2-335},
doi = {10.1364/JOSAA.17.000335},

}

@article{qplate_matrix0,
  title = {Optical Spin-to-Orbital Angular Momentum Conversion in Inhomogeneous Anisotropic Media},
  author = {Marrucci, L. and Manzo, C. and Paparo, D.},
  journal = {Phys. Rev. Lett.},
  volume = {96},
  issue = {16},
  pages = {163905},
  numpages = {4},
  year = {2006},
  month = {04},
  publisher = {American Physical Society},
  doi = {10.1103/PhysRevLett.96.163905},
  url = {https://link.aps.org/doi/10.1103/PhysRevLett.96.163905}
}

@article{Nivas2017,
author = {JJ Nivas, Jijil and Cardano, Filippo and Song, Zhenming and Rubano, Andrea and Fittipaldi, Rosalba and Vecchione, Antonio and Paparo, Domenico and Marrucci, Lorenzo and Bruzzese, Riccardo and Amoruso, Salvatore},
year = {2017},
month = {02},
pages = {42142},
title = {Surface Structuring with Polarization-Singular Femtosecond Laser Beams Generated by a q-plate},
volume = {7},
journal = {Scientific Reports},
doi = {10.1038/srep42142}
}

@article{walborn2010spatial,
  title={Spatial correlations in parametric down-conversion},
  author={Walborn, Stephen P and Monken, CH and P{\'a}dua, S and Ribeiro, PH Souto},
  journal={Phys. Rep.},
  volume={495},
  number={4-5},
  pages={87--139},
  year={2010},
  publisher={Elsevier}
}

@article{fuenzalida2022resolution,
  title={Resolution of quantum imaging with undetected photons},
  author={Fuenzalida, Jorge and Hochrainer, Armin and Lemos, Gabriela Barreto and Ortega, Evelyn A and Lapkiewicz, Radek and Lahiri, Mayukh and Zeilinger, Anton},
  journal={Quantum},
  volume={6},
  pages={646},
  year={2022},
  publisher={Verein zur F{\"o}rderung des Open Access Publizierens in den Quantenwissenschaften}
}

@article{Yao:11,
author = {Alison M. Yao and Miles J. Padgett},
journal = {Adv. Opt. Photon.},
number = {2},
pages = {161--204},
publisher = {Optica Publishing Group},
title = {Orbital angular momentum: origins, behavior and applications},
volume = {3},
month = {06},
year = {2011},
url = {https://opg.optica.org/aop/abstract.cfm?URI=aop-3-2-161},
doi = {10.1364/AOP.3.000161}
}

@article{yamaguchi1997phase,
  title={Phase-shifting digital holography},
  author={Yamaguchi, Ichirou and Zhang, Tong},
  journal={Opt. Lett.},
  volume={22},
  number={16},
  pages={1268--1270},
  year={1997},
  publisher={Optical Society of America}
}

@article{topfer2022quantum,
  title={Quantum holography with undetected light},
  author={T{\"o}pfer, Sebastian and Gilaberte Basset, Marta and Fuenzalida, Jorge and Steinlechner, Fabian and Torres, Juan P and Gr{\"a}fe, Markus},
  journal={Sci. Adv.},
  volume={8},
  number={2},
  pages={eabl4301},
  year={2022},
  publisher={American Association for the Advancement of Science}
}

@article{pearce2024single,
  title={Single-frame transmission and phase imaging using off-axis holography with undetected photons},
  author={Pearce, Emma and Wolley, Osian and Mekhail, Simon P and Gregory, Thomas and Gemmell, Nathan R and Oulton, Rupert F and Clark, Alex S and Phillips, Chris C and Padgett, Miles J},
  journal={Sci. Rep.},
  volume={14},
  number={1},
  pages={16008},
  year={2024},
  publisher={Nature Publishing Group UK London}
}

@article{fuenzalida2024nonlinear,
  title={Nonlinear Interferometry: A New Approach for Imaging and Sensing},
  author={Fuenzalida, Jorge and Giese, Enno and Gr{\"a}fe, Markus},
  journal={Adv. Quantum Technol.},
  pages={2300353},
  year={2024},
  publisher={Wiley Online Library}
}

@article{fuenzalida2023experimental,
  title={Experimental quantum imaging distillation with undetected light},
  author={Fuenzalida, Jorge and Gilaberte Basset, Marta and T{\"o}pfer, Sebastian and Torres, Juan P and Gr{\"a}fe, Markus},
  journal={Sci. Adv.},
  volume={9},
  number={35},
  pages={eadg9573},
  year={2023},
  publisher={American Association for the Advancement of Science},
url={https://doi/10.1126/sciadv.adg9573}
}

@article{leon2024off,
  title={Off-axis holographic imaging with undetected light},
  author={Le{\'o}n-Torres, Josu{\'e} R and Krajini{\'c}, Filip and Kumar, Mohit and Gilaberte Basset, Marta and Setzpfandt, Frank and Gili, Valerio Flavio and Jelenkovi{\'c}, Branislav and Gr{\"a}fe, Markus},
  journal={Opt. Express},
  volume={32},
  number={20},
  pages={35449--35461},
  year={2024},
  publisher={Optica Publishing Group}
}

@article{topfer2025synthetic,
  title={Synthetic quantum holography with undetected light},
  author={T{\"o}pfer, Sebastian and Tovar-Perez, Sergio and Le{\'o}n Torres, Josu{\'e} R and Derr, Daniel and Giese, Enno and Fuenzalida, Jorge and Gr{\"a}fe, Markus},
  journal={Optica Quantum},
  volume={3},
  number={2},
  pages={129--136},
  year={2025},
  publisher={Optica Publishing Group}
}

@article{Szadowiak:2021lpu,
    author = "Szadowiak, Wiktor and Kundu, Sanjukta and Szuniewicz, Jerzy and Lapkiewicz, Radek",
    title = "{Self-referenced hologram of a single photon beam}",
    eprint = "2006.02580",
    archivePrefix = "arXiv",
    primaryClass = "quant-ph",
    doi = "10.22331/q-2021-08-03-516",
    journal = "Quantum",
    volume = "5",
    pages = "516",
    year = "2021"
}

@article{Sanchez_Ortiga:2014AO,
author = {Sánchez-Ortiga, Emilio and Doblas, Ana and Saavedra, Genaro and Martinez-Corral, Manuel and Garcia-Sucerquia, Jorge},
year = {2014},
month = {03},
pages = {2058-2066},
title = {Off-axis digital holographic microscopy: practical design parameters for operating at diffraction limit},
volume = {53},
journal = {Applied Optics},
doi = {10.1364/AO.53.002058},
}

@article{Verrier:11,
author = {Nicolas Verrier and Michael Atlan},
journal = {Appl. Opt.},
number = {34},
pages = {H136--H146},
publisher = {Optica Publishing Group},
title = {Off-axis digital hologram reconstruction: some practical considerations},
volume = {50},
month = {12},
year = {2011},
url = {https://opg.optica.org/ao/abstract.cfm?URI=ao-50-34-H136},
doi = {10.1364/AO.50.00H136},

}

@article{Cuche:00,
author = {Etienne Cuche and Pierre Marquet and Christian Depeursinge},
journal = {Appl. Opt.},
number = {23},
pages = {4070--4075},
publisher = {Optica Publishing Group},
title = {Spatial filtering for zero-order and twin-image elimination in digital off-axis holography},
volume = {39},
month = {08},
year = {2000},
url = {https://opg.optica.org/ao/abstract.cfm?URI=ao-39-23-4070},
doi = {10.1364/AO.39.004070},

}

@article{Brambila:25,
author = {Emma Brambila and Raphael Guitter and Ren\'{e} Sondenheimer and Markus Gr\"{a}fe and Hugo Defienne},
journal = {Opt. Lett.},
number = {15},
pages = {4854--4857},
publisher = {Optica Publishing Group},
title = {Certifying spatial entanglement between non-degenerate photon pairs with a camera},
volume = {50},
month = {08},
year = {2025},
url = {https://opg.optica.org/ol/abstract.cfm?URI=ol-50-15-4854},
doi = {10.1364/OL.569692},
}

\end{document}